\documentclass[onecolumn]{aastex6}

\shorttitle{Sub-mm variability of EC~53}
\shortauthors{Yoo et al.}

\begin{document}

\title{The JCMT Transient Survey: Detection of sub-mm variability in a Class I protostar EC 53 in Serpens Main}

\author{Hyunju Yoo\altaffilmark{1,2}, Jeong-Eun Lee\altaffilmark{2,3}, 
            Steve Mairs\altaffilmark{4,5}, Doug Johnstone\altaffilmark{4,5}, Gregory J. Herczeg\altaffilmark{6}, 
            Sung-ju Kang\altaffilmark{7}, Miju Kang\altaffilmark{7}, Jungyeon Cho\altaffilmark{1} and The JCMT Transient Team}

\altaffiltext{1}{Department of Astronomy and Space Science, Chungnam National University, 99 Daehak-ro, Yuseong-gu, Daejeon 34134, Republic of Korea}
\altaffiltext{2}{School of Space Research, Kyung Hee University, 1732, Deogyeong-Daero, Giheung-gu Yongin-shi, Gyunggi-do 17104, Republic of Korea}
\altaffiltext{3}{jeongeun.lee@khu.ac.kr}
\altaffiltext{4}{NRC Herzberg Astronomy and Astrophysics, 5071 West Saanich Rd, Victoria, BC, V9E 2E7, Canada}
\altaffiltext{5}{Department of Physics and Astronomy, University of Victoria, Victoria, BC, V8P 1A1, Canada}
\altaffiltext{6}{Kavli Institute for Astronomy and Astrophysics, Peking University, Yiheyuan 5, Haidian Qu, 100871 Beijing, People's Republic of China}
\altaffiltext{7}{Korea Astronomy and Space Science Institute, 776 Daedeokdae-ro, Yuseong-gu, Daejeon 34055, Republic of Korea}



\begin{abstract}
During the protostellar phase of stellar evolution, accretion onto the
star is expected to be variable, but this suspected variability has been
difficult to detect because protostars are deeply embedded.  In this
paper, we describe a sub-mm luminosity burst of the Class I
protostar EC~53 in 
Serpens Main, the first variable found during our dedicated
JCMT/SCUBA-2 monitoring program of eight nearby star-forming regions.   
EC~53 remained
quiescent for the first 6 months of our survey, from February to
August 2016.  The sub-mm emission began to brighten in September 2016,
reached a peak brightness of $1.5$ times the faint state, and has been
decaying slowly since February 2017.
The change in sub-mm brightness is interpreted as dust heating in the
envelope, generated by a luminosity increase of the protostar of a
factor of $\ge 4$.   
The 850~$\mu$m lightcurve resembles the historical $K$-band
lightcurve, which varies by a factor of $\sim 6$ with a 543 period
and is interpreted as accretion variability excited by interactions
between the accretion disk and a close binary system. 
The predictable detections of accretion variability observed at both near-infrared and sub-mm
wavelengths make the system a unique test-bed, enabling us to capture the
moment of the accretion burst and to study the consequences of the outburst
on the protostellar disk and envelope.  
\end{abstract}






\section{Introduction} \label{sec:intro}

A fundamental question of star formation is how stars gain their
mass. In our current paradigm of low mass star formation, young stars
form via the gravitational collapse of cold, dense molecular
cores. Protostellar disks serve as temporary mass reservoirs,
accreting material from the envelope and channeling that material onto
the protostar. The mass accretion rate in the early stages of stellar
assembly determines the physical characteristics of the protostar
(e.g. mass and luminosity).  In historical models of protostellar
collapse, the star grows at a constant rate \citep{shu77,masunaga00}.
However, the distribution of luminosities of young stellar objects (YSOs) is
an order of magnitude fainter than expected from the standard model of
dynamical collapse, in what has become known as the {\it Luminosity Problem} \citep{kenyon90,dunham10}.
Episodic accretion, which has quiescent-accretion phases
interspersed with burst-accretion phases, is a
promising solution for the luminosity problem (see discussion in
\citealt{fischer17} and references therein). 

The more evolved Class II YSOs (classical T Tauri stars) provide
support for the importance of accretion variability 
objects because these star+disk systems are optically visible \citep[see
  reviewby][]{hartmann16}.
Optical and near-IR monitoring of accretion demonstrates that the
variability spans a wide range of timescales and 
amplitudes.  Large, but rare eruptions correspond to increases in accretion rate
by factors of $10^2-10^4$ and can last months-to-decades (see compilation by \citealt{audard14} and frequency analysis by \citealt{hillenbrand15}).
Several stars with known eruptions have retained envelopes and are considered Class I
YSOs, including 
V1647 Ori \citep{reipurth04,andrews04}, 
OO~Ser \citep{hodapp96,hodapp12,kospal07}, 
V2492 Cyg \citep{covey11},
V2775~Ori \citep{fischer12}, and
V900~Mon \citep{reipurth12}.
Although large changes in accretion rate are only rarely detected in
the optical and near-IR, accretion flickers of a factor of a few occur
routinely \citep[e.g.][]{costigan14,venuti14,cody17}. 

During the earlier, main phase of stellar
assembly, the protostar is surrounded by a dense envelope, so
detecting accretion variability is challenging at
optical/near-IR wavelengths.
In these systems, most of the accretion energy eventually emerges at
far-IR and sub-mm wavelengths \citep{johnstone13}, where objects have
not been previously monitored because of limitations in flux
calibration.  The only deeply embedded 
protostars with a detected change in sub-mm
brightness are HOPS~383, which brightened by a factor of 2 in the
sub-mm (350~$\mu m$ and
450~$\mu m$) compared to a factor of 35 at 24 $\mu$m \citep{safron15},
and the massive protostellar system NGC\,6334\,I-MM1
\citep{hunter17}. However, during this early life  
of a protostar, the accretion variability may be enhanced:  accretion
rates, disk masses, and envelope masses are all higher, and any
stellar or planetary-mass companions may be in dynamically-unstable
orbits that perturb any disk in the system.

 The dynamical mechanisms that trigger episodes of accretion bursts
 and flickers are still unclear.
Thermal instability in inner disk \citep{bell94}, gravitational instability
\citep[e.g.][]{vorobyov10}, and the combined effect of magnetorotational instability and gravitational
instability \citep{zhu09,bae14} 
are possible causes of
long-lasting and large eruptions, while short bursts are likely
magnetospheric \citep[e.g.][]{dangelo10}.  Accretion bursts may also be tidally induced by
gravitational interaction with the companion star in a binary system \citep{bonnell92}
or caused by an encounter with a passing
star \citep{forgan10}. Furthermore, a massive protoplanet also
can trigger the outburst \citep{lodato04}. The triggering
mechanisms may depend on the physical and dynamical conditions in the
system, which would lead to different time evolutions of the accretion
luminosity \citep{audard14} and different disk
structures. 

In this paper, we present a new detection of a sub-mm variable source
associated with the protostar EC 53 \citep[also known as V371 Ser and
Ser SMM5;][]{eiroa92,casali93}, located in the
active star-forming region Serpens Main \citep{eiroa08} at a
distance of $436\pm9.2$ pc \citep{ortiz17ser}.  The sub-mm variability
is identified in monthly James Clerk Maxwell Telescope (JCMT) SCUBA-2 
monitoring as part of the JCMT-Transient Survey \citep{herczeg17}
and is the 
first sub-mm protostellar variable identified in any dedicated sub-mm monitoring program.  
The sub-mm lightcurve matches well
the previously-known 543-day periodic variability in the $K$-band \citep{hodapp99,hodapp12}.

EC~53 was first identified in near-IR imaging as a protostar with associated nebulosity in the near-IR
\citep{eiroa92} and strong sub-mm emission, indicating the presence of
an envelope \citep{casali93}.  The bolometric temperature of 130--240
K \citep{Evans09,dunham15}, the spectral slope of $\sim 0.7-1$
\citep{enoch09,gutermuth09,dunham15}, and the ratio of sub-mm to
bolometric luminosity of $\sim 0.3$ \citep{dionatos10} are all
consistent with the rough definition of a Class I YSO.
Outflows from the protostar have been detected in sub-mm CO rotational
emission and in near-IR H$_2$ rovibrational
emission \citep{dionatos10,hodapp12}.  The
bolometric luminosity has been measured to be $1.7-4.8$~L$_\odot$
\citep{enoch09,Evans09,dunham15}; the central source is also X-ray
bright \citep{giardino07}.  High-resolution imaging has resolved the
protostar into an $0\farcs3$ (129 AU) binary, where the secondary is
very faint and either very low-mass, more deeply embedded than EC 53A,
or located behind the EC 53 envelope \citep{hodapp12}.  Variability
was first detected in the near-IR by \citet{hodapp96} and
\citet{horrobin97}.   Continued monitoring later revealed a
periodicity in the $K$-band brightness that is likely attributed to an
unresolved binarity of the primary \citep{hodapp12}.  The near-IR
spectrum shows deep CO overtone band absorption, which is typical of
the  hot disks of FUor outbursts  \citep{doppmann05}. 

The outline of this paper is as follows: We start by explaining the
sub-mm monitoring observations using JCMT/SCUBA-2 in Section
\ref{sec:obs}. In Section \ref{sec:data}, we describe the data
reduction, including flux calibration (Section \ref{subsec:cal}) and
source identification (Section
\ref{subsec:sour}). We then measure the peak brightness of the bright sub-mm
sources in each epoch and identify the significant variability of peak brightness from EC 53 in Section \ref{sec:result}. In Section
\ref{sec:discussion}, we compare our 850~$\mu m$ lightcurve to
previously reported K-band lightcurves \citep{hodapp99,hodapp12}, 
discuss binarity as a potential cause of the variability, and describe 
the potential scientific utility that predictable changes in the accretion rate provide.  Finally, we summarize our results in Section \ref{sec:summary}. 

\vskip 0.5cm
\section{Observations} \label{sec:obs}

In the JCMT Transient Survey \citep{herczeg17}, we are
monitoring eight nearby active star-forming 
regions for three years at 450 and
850~$\mu$m with JCMT/SCUBA-2 \citep[Submillimetre Common User
Bolometer Array 2,][]{holland13}. The main purpose of this project is
to quantify the accretion variability of deeply embedded protostars from
monthly monitoring of the sub-mm continuum emission.  As part of this
project, we observed the Serpens Main 12 times between 2 February 2016 and 17 April
2017 (see Table \ref{tab:obs}).  The cadence of our imaging was once
every $\sim 4$ weeks from 
2 February 2016 to 20 March 2016, and increased to once every $\sim 2$ weeks from 
April 2017 to better resolve the decay of EC 53. 
Serpens Main was not
observed from October 2016 through January 2017 because the field
was too close to the Sun and not visible.

Our SCUBA-2 imaging consists of a pong 1800$\arcsec$ mapping mode that covers a circular
region with a $30^\prime$ diameter \citep{kackley10} centered at R.A.=18$^h$29$^m$49$^s$,
Dec.=+01$\degr$15$\arcmin$20$\arcsec$. Integration times are
determined based on the measured precipitable water vapor in the atmosphere to reach a
sensitivity of $\sim 12$ mJy~beam$^{-1}$ at 850 $\mu$m.
The effective beam
size of SCUBA-2 at 850~$\mu m$ is 14.6$''$ \citep{dempsey13}. 
The observation on 17 March 2016
is excluded from our analysis because of pointing offset issues in the observing run.

Our final dataset in this paper consists of 11 epochs of 850 $\mu$m photometry.
Although SCUBA-2 images simultaneously at 450 and 850 $\mu$m, in this 
paper we concentrate on only the 850 $\mu$m emission because our flux
calibration techniques have only been developed for those images.
We see hints of variability in the 450 $\mu$m, but the 450 $\mu$m maps are noisier and 
suffer more from changes in the beam profile and varying atmospheric opacity. 
Therefore, we leave the more detailed analysis for later.

\begin{deluxetable}{ccccc}
\tablecaption{A summary of observations at 850~$\mu m$. 
\label{tab:obs}}
\tablewidth{0pt}
\tabletypesize{\small}
\tablehead{
\colhead{Date} & \colhead{Julian day} & \colhead { $\tau_{225}$$^a$} & \colhead {Zenith} & \colhead{Noise} \\
\colhead{(yyyy-mm-dd)} & \colhead{ } & \colhead{} & \colhead{Opacity$^b$} & \colhead{(mJy~beam$^{-1}$)}
}
\startdata
2016-02-02 &	     2457420.72	&	0.08	&	0.09	& 13\\
2016-02-23 &  	     2457441.68	&	0.04	&	0.05 & 12\\ 
2016-03-17$^c$ & 2457493.54 & 	0.03	&	0.04	& 13 \\
2016-04-15 &	     2457493.54	& 	0.03 	&	0.04 & 11\\
2016-05-21 &	     2457529.39	& 	0.06 	&	0.08 & 15\\
2016-07-22 &	     2457591.44	& 	0.08 	&	0.10 & 13\\
2016-08-27 &	     2457627.23	& 	0.08	&	0.09 & 12\\
2016-09-29 &	     2457660.22	& 	0.09	&	0.10 & 13\\
2017-02-22 &	     2457806.66	& 	0.06 	&	0.10 & 11\\
2017-03-20 &	     2457832.58	& 	0.05 	&	0.07 & 12\\
2017-04-03 &	     2457846.56	& 	0.06 	&	0.06 & 11\\
2017-04-17 &        2457860.54	& 	0.04 	&	0.06 & 10\\
\enddata
\tablenotetext{a}{The average 225 GHz zenith opacity measured during observation.}
\tablenotetext{b}{The average zenith opacity measured with the water
  vapour monitor at the JCMT during observation.} 
\tablenotetext{c}{The image of this epoch shows elongated structures of
  structures due to pointing-drift during observation and is omitted
  for data analysis.} 
\end{deluxetable}

\begin{figure*}[!t]
\centering
\includegraphics[width=.48\textwidth]{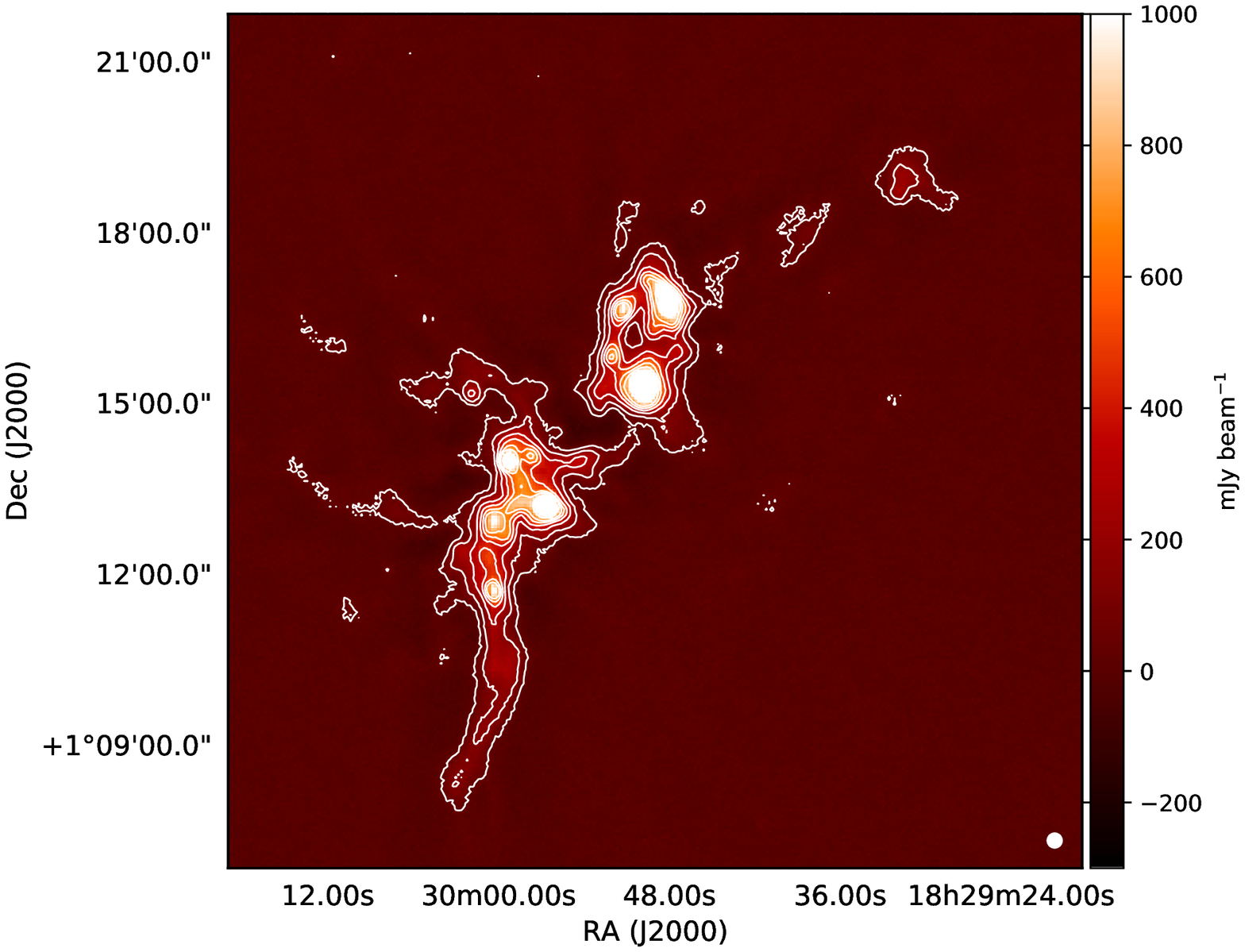}\hfill
\includegraphics[width=.48\textwidth]{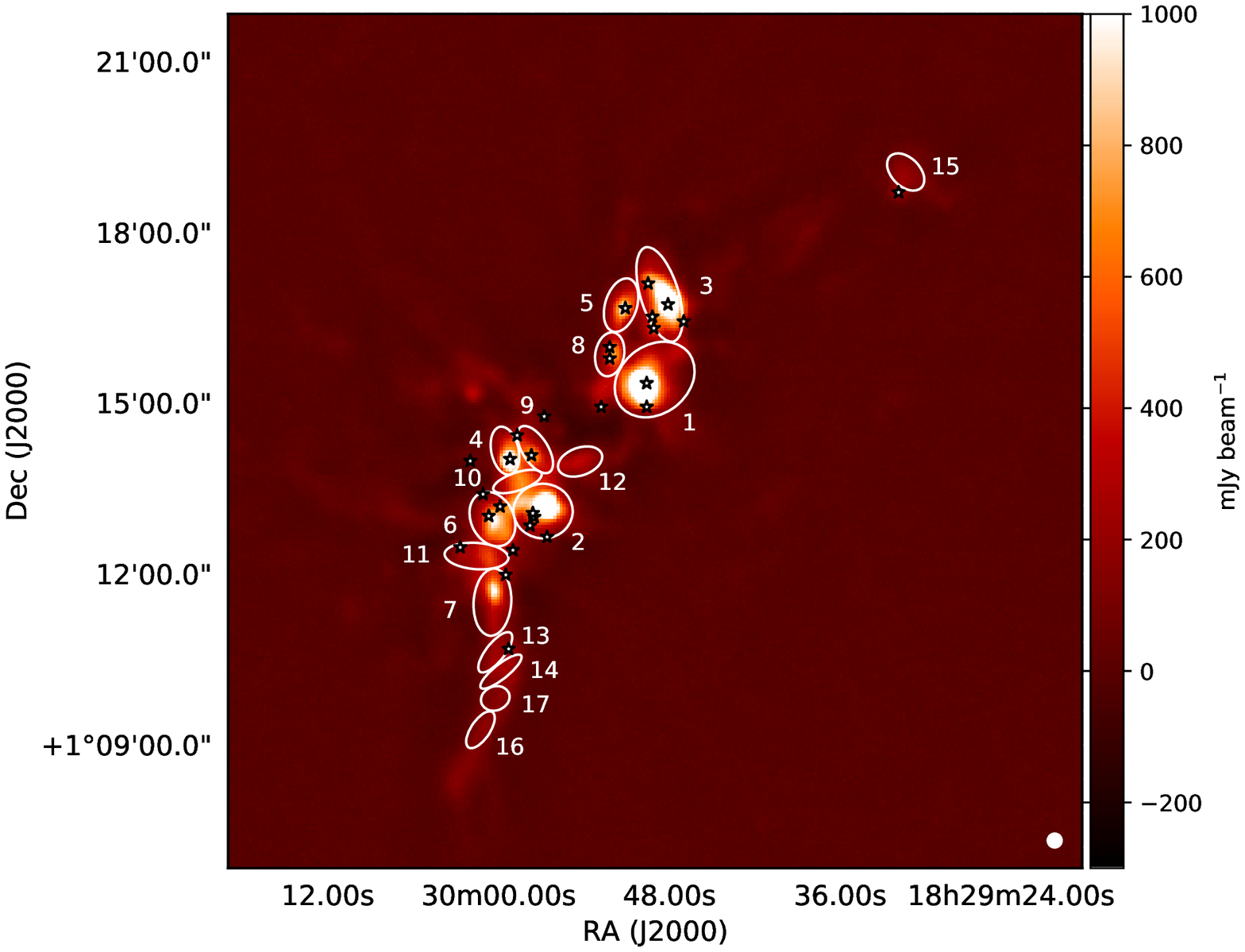}
\caption{The 850~$\mu m$ map of Serpens Main. Left panel: The
  contour levels in the map are 10, 40, 80, 120, 160, 200, and 300~$\sigma$
  with the mean rms noise of 3.6~mJy~beam$^{-1}$ (1$\sigma$). 
  Right panel: White ellipses show the locations of 17 clumps identified within
  the region and black stars mark the positions of protostars listed in
  \citet{dunham15}. The JCMT beam at 850~$\mu m$ (14.6$''$) is
  presented as a white circle at the bottom right corner of both
  panels. 
\label{fig:SM_coadd}}
\end{figure*}

\vskip 0.5cm
\section{Data reduction} \label{sec:data}

\subsection{Flux calibration} \label{subsec:cal}

In this section, we provide a brief overview of the data reduction and flux
calibration of the SCUBA-2 observations (see \citealt{mairs15} and \citealt{mairs17} 
for further details).  
Each Transient Survey SCUBA-2 observation is performed by continuously scanning across the sky 
to fill in a circular region 30$\arcmin$ in diameter. 
The dominant noise sources in the map are the signal
received from the atmosphere and the instrument, both of which vary
with time and are 
uncorrelated with location within the map. Meanwhile, the scan pattern
assures that the astronomical signal, which is dependent on map position, is 
observed from multiple position angles.
In order to produce the most robust map, the sources of noise are
modeled and removed from the final image using the iterative map-making
process MAKEMAP \citep{chapin13}, found in the SMURF package
\citep{jenness13} in the STARLINK software \citep{currie14}. MAKEMAP applies the
flat-field correction, removes the atmospheric signal (common mode) by
filtering out large scale structures, applies the water vapor
extinction correction, and performs additional high pass filtering
based on several user supplied parameters.  Regions with strong
emission are identified and masked for a subsequent reduction (see
\citealt{chapin13} and \citealt{mairs15} for further information).

After recovering the astronomical signal in the individual maps, the
pointing offsets between each observation are measured and the
observations are then re-reduced a final time to place all images on the same pixel
grid.  
The nominal JCMT pointing uncertainty of 2--6$\arcsec$ is
improved to $\sim 0\farcs5$ by measuring the centroid of bright, compact sources
in the field and re-centering the image \citep[for details
see][]{mairs17}.  The final map has a pixel scale of
$3^{\prime\prime}$.

\citet{mairs17} tested four different data reduction methods to
optimize flux calibration of compact sources for our observations in the JCMT-Transient Survey,
altering the amount of large-scale structure removed from the final
image and whether or not an external mask was applied. We adopt the
reductions that filter out emission on scales larger than 200$\arcsec$
\citep[R3 reduction in][]{mairs17} and masks regions with strong
emission, though all reductions tested produce consistent results for
EC 53.

Finally, the relative flux in each source is measured across all
epochs by first identifying a set of stable `calibrator' sources within the map 
and then using these sources to bring all epochs into agreement \citep{mairs17}.
The default absolute flux calibration uncertainty for SCUBA-2 images is 
5--10\% (see \citealt{dempsey13} and \citealt{mairs17}).  Calibrating
fluxes relative to the fluxes of bright, stable, compact sources improves the
flux calibration uncertainty to 2--3\%.  For Serpens Main, our
flux calibration is based on a set of five sources (Sources 2, 3, 4, 6,
and 7 in Table \ref{tab:indiv}), which are all brighter than 1~Jy
beam$^{-1}$, have radii smaller than
10$''$, and have a stable normalised peak brightness with
respect to one another, as measured in the JCMT-Transient images
obtained between February 2016 --  March 2017 \citep[see also \S 4 and][]{mairs17}. 
  
\tabletypesize{\scriptsize}
\begin{deluxetable*}{lccc|ccccccccccc|ccc}
\tablecaption{The peak brightness of individual sources in 850~$\mu m$. 
\label{tab:indiv}}
\tablewidth{0pt} \setlength{\tabcolsep}{0.035in}
\tablehead{
\colhead{} & \colhead{Date$^a$} & \colhead{}  &\colhead{} & \colhead{16-02-02} & \colhead{16-02-23} & \colhead{16-04-15} & \colhead{16-05-21} & \colhead{16-07-22} & \colhead{16-08-27} & \colhead{16-09-29} & \colhead{17-02-22} & \colhead{17-03-20} & \colhead{17-04-03} & \colhead{17-04-17} & \colhead{} & \colhead{} & \colhead{}\\
\hline
\colhead{} & \colhead{MJD$^b$} & \colhead{}&\colhead{} &  \colhead{57421} & \colhead{57442} & \colhead{57494} & \colhead{57529} & \colhead{57591} & \colhead{57627} & \colhead{57660} & \colhead{57807} & \colhead{57833} & \colhead{57847} & \colhead{57861}& \colhead{} & \colhead{} & \colhead{}\\
\hline
\colhead{ID$^c$}  & R.A.$^d$ & Dec.$^d$ & \multicolumn{12}{c}{Peak brightness (mJy~beam$^{-1}$)} & Mean $^e$ & S.D.$^e$ & F.V.$^f$
}
\startdata
\hline
1$^c$ & 18 29 49.0  & +01 15 24.8 & & 6380 & 6440 & 6780 & 6300 & 6670 & 6800 & 6920 & 6760 & 6530 & 6460 & 6550 & 6600 & 190 & 0.029\\
2$^c$$^g$ & 18 29 56.9  & +01 13 06.2 & & 2870 & 2890 & 3000 & 2990 & 2940 & 2930 & 3020 & 2970 & 2880 & 2890 & 2900 & 2930 & 50 & 0.017 \\
3$^c$$^g$ & 18 29 48.7  & +01 16 54.5& & 1970 & 1990 & 1940 & 1960 & 1910 & 1870 & 1960 & 1970 & 2000 & 1980 & 1970 & 1960 & 40 & 0.018 \\
4$^c$$^g$ & 18 29 59.6  & +01 14 09.8 & & 1690 & 1720 & 1720 & 1690 & 1700 & 1690 & 1710 & 1700 & 1660 & 1670 & 1660 & 1690 & 20 & 0.013\\ 
5$^c$ & 18 29 51.4  & +01 16 43.2 & & 1000 &  930 &  950 &  940 &  980 &  960 & 1110 & 1290 & 1250 & 1150 & 1220 & 1070 & 130 & 0.122\\
6$^c$$^g$ & 18 30 00.4  & +01 12 57.8 & & 1040 & 1060 & 1030 & 1030 & 1080 & 1040 & 1050 & 1040 & 1070 & 1070 & 1070 & 1050 & 20 & 0.017\\
7$^c$$^g$ & 18 30 00.4  & +01 11 30.2 & & 1090 & 1050 & 1080 & 1040 & 1070 & 1030 & 1100 & 1080 & 1007 & 1030 & 1040 & 1060 & 30 & 0.023\\
8$^c$ & 18 29 52.2  & +01 15 51.4 & & 750 &  770 &  760 &  800 &  760 &  740 &  850 &  800 &  810 &  770 &  780 & 780 & 20 & 0.038\\
9$^c$ & 18 29 57.4  & +01 14 11.0 & & 760 &  790 &  760 &  770 &  800 &  810 &  750 &  780 &  740 &  760 &  770 & 770 & 30 & 0.026\\
10 & 18 29 58.7  & +01 13 37.4 & &  750 &  770 &  700 &  790 &  790 &  770 &  780 &  750 &  780 &  800 &  720 & 760 & 30 & 0.039\\
11$^c$ & 18 30 01.6  & +01 12 18.8 & &  610 &  550 &  550 &  580 &  550 &  520 &  540 &  560 &  560 &  550 &  550 & 560 & 20 & 0.039\\
12 & 18 29 54.3  & +01 13 58.4 & &  390 &  370 &  380 &  370 &  380 &  410 &  410 &  400 &  370 &  380 &  380 & 390 & 10 & 0.035\\
13$^c$ & 18 30 00.2  & +01 10 37.4 & &  350 &  320 &  310 &  340 &  350 &  290 &  340 &  300 &  300 &  310 &  270 & 320 & 20 & 0.074\\ 
14 & 18 29 59.8 & +01 10 17.0 & &  310 &  330 &  300 &  320 &  310 &  300 &  290 &  300 &  320 &  300 &  300 & 310 & 10 & 0.037\\
15 & 18 29 31.4  & +01 19 03.5 & &  270 &  250 &  270 &  310 &  270 &  270 &  260 &  300 &  250 &  240 &  240 & 270 & 20 & 0.084\\
16 & 18 30 01.9  & +01 09 15.8 & &  250 &  200 &  240 &  240 &  250 &  240 &  220 &  220 &  220 &  200 &  210 & 230 & 20 & 0.077\\
17 & 18 30 00.2  & +01 09 48.8 & &  230 &  220 &  220 &  240 &  260 &  210 &  200 &  240 &  260 &  240 &  200 & 230 & 20 & 0.091\\
\enddata
\tablenotetext{a}{Date in the format of YY-MM-DD.}
\tablenotetext{b}{Modified Julian Date of observation.}
\tablenotetext{c}{Clumps that contain at least one protostar.}
\tablenotetext{d}{Coordinates of the center of the regions in the right panel of Figure \ref{fig:SM_coadd}.}
\tablenotetext{e}{Mean peak brightness and the standard deviation of individual source calculated over 11 epochs.}
\tablenotetext{f}{Fractional variance is obtained from standard deviation divided by the mean peak brightness for each source.}
\tablenotetext{g}{Calibrators.}
\end{deluxetable*}


\subsection{Source identification} \label{subsec:sour}

We coadded all 11 epochs to create a deep map and identify faint sources.  The
average rms noise of $\sim$12~mJy~beam$^{-1}$ in each individual map
(as measured from the regions where no emission source
exists) improves to $\sim$3.6~mJy~beam$^{-1}$ in the final coadded map.  
The right panel of
Figure \ref{fig:SM_coadd} shows that several groups of compact sources
are distributed along the main filament.   

In order to identify sub-mm sources, we used the ClumpFind algorithm
\citep{williams94} in the STARLINK package \citep{currie14}. Sources identified by ClumpFind that 
are smaller than the beam size are 
excluded from further analysis\footnote{Sources are excluded if the 
clump contains fewer than 20 pixels, which 
corresponds to a radius of $\sim$7\farcs5.}.
We find 17 robust clumps with intensities brighter than 10 times the rms
noise (see the right panel of Figure \ref{fig:SM_coadd}).   
Table \ref{tab:indiv} lists the peak brightness of the 17 identified clumps measured from each
epoch map and the mean peak brightness of individual clumps in 11 epochs.  
The peak brightness of each clump is used to produce its lightcurve. 
 An independent analysis using the GaussClumps algorithm \citep{stutzki1990} shows trends in the lightcurves that are consistent with the results described here.

\vskip 0.5cm
\section{Results} \label{sec:result}

\begin{figure*}[!tp]
\centering
\includegraphics[width=.48\textwidth]{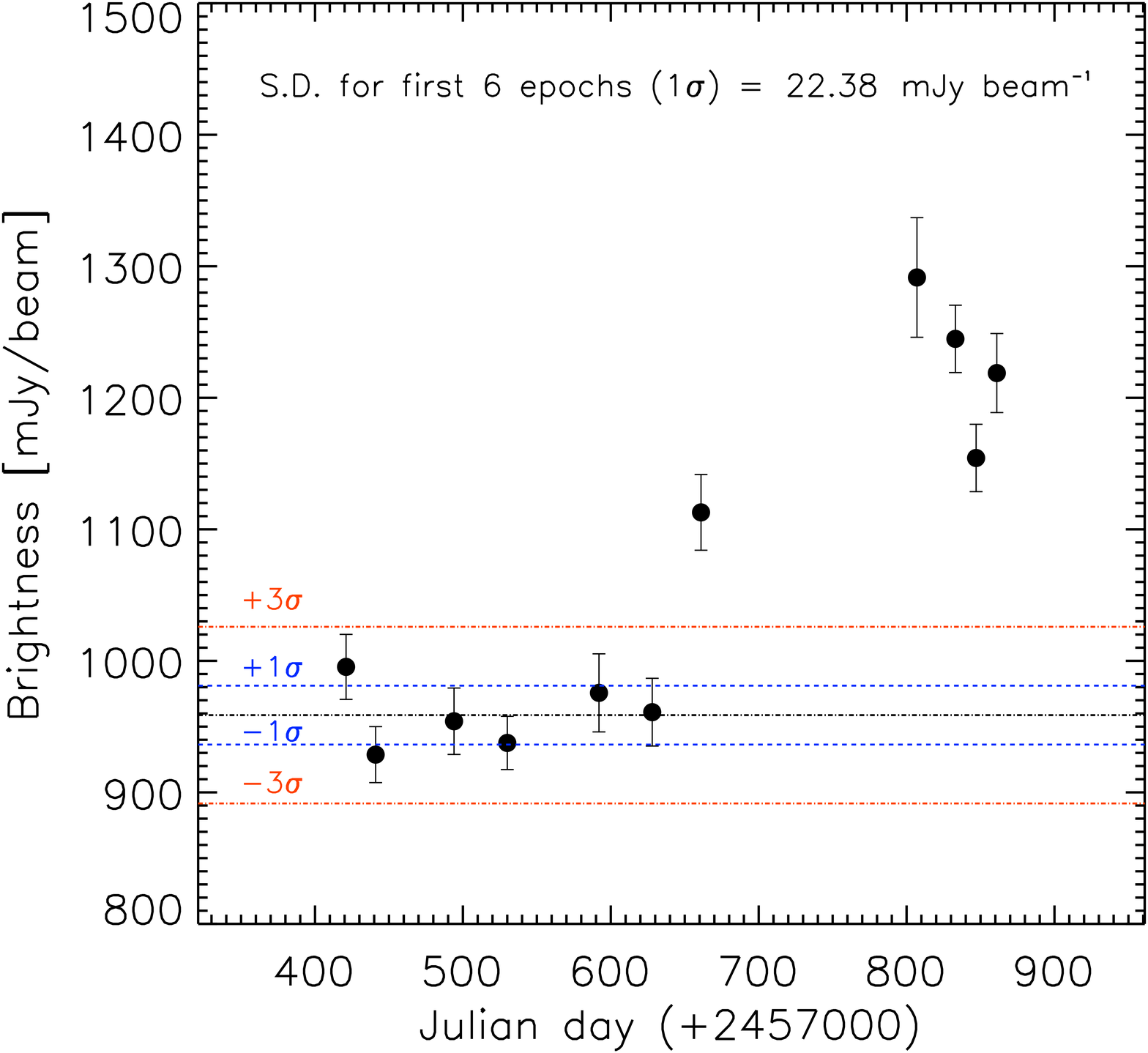}\hfill
\includegraphics[width=.48\textwidth]{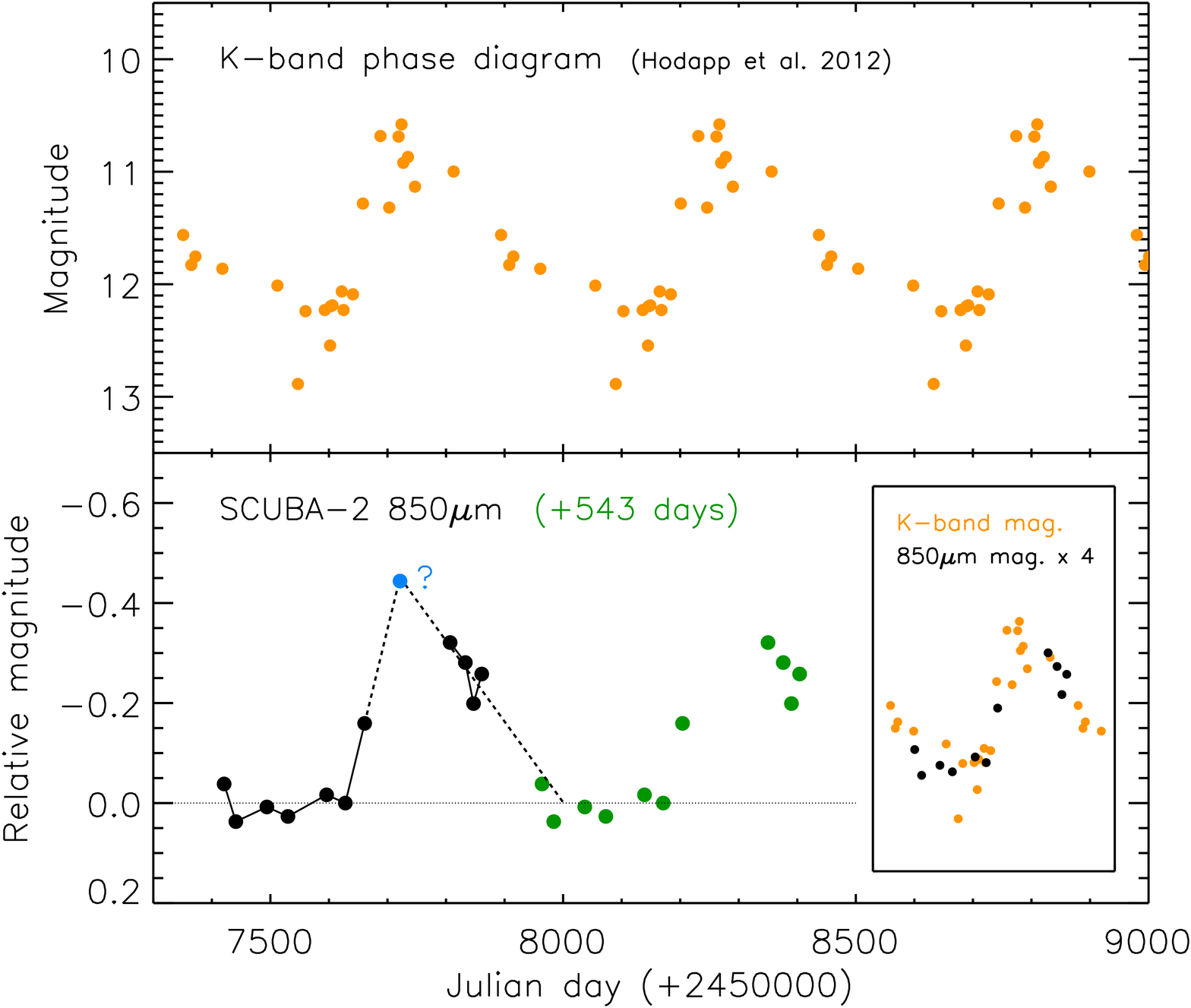}
\caption{Left panel: The time variation of peak brightness of Source~5 (EC~53) at
  850~$\mu m$. The blue and red horizontal lines indicate
  $\pm$1$\sigma$ and $\pm$3$\sigma$ levels (1$\sigma$ $\sim$
  22.4~mJy~beam$^{-1}$ is the standard deviation of peak brightness for
  the first 6 epochs) relative to the mean peak brightness over the first 6
  epochs. Upper right panel: Replicated K-band lightcurve of EC~53
  from the phase diagram in Figure 5 of \citet{hodapp12} with a period of
  543 days. Lower right panel: Relative magnitude with respect to the
  6th epoch (zero magnitude) observed on August 2016. The black circles
  are the JCMT/SCUBA-2 850~$\mu m$ data points observed from February 2016
  to April 2017 and the green circles are replicated 850~$\mu m$ data
  assuming the sub-mm lightcurve has the same 543 day period as in the
  K-band. The blue circle indicates the expected maximum
  brightness, which is calculated by comparing the brightening in the 6th and 7th epochs and 
  the decay of the last four epochs. 
  The inset shows the 850 $\mu$m lightcurve, whose magnitude is amplified by a factor of 4, 
  on top of the K-band lightcurve.
\label{fig:lc}}
\end{figure*}

We monitored sub-mm emission from the 17 brightest clumps in the
Serpens Main star-forming region with JCMT/SCUBA-2 at 11 epochs
between 2 February 2016 to 17 April 2017 (Table \ref{tab:indiv}).   
Most sources
appear to have stable peak brightness throughout this observational period.
Source 5, coincident with the near-IR YSO EC 53 (hereafter
referred to as EC 53), began to brighten in September, reached a peak
when Serpens was behind the Sun and not visible, and fades
slightly from February 2017 to April 2017 
(Figure \ref{fig:lc}). 
The peak brightness during the first six epochs is stable at 960
mJy~beam$^{-1}$ with a standard deviation of 22 mJy~beam$^{-1}$,  
consistent with the expected level of random measurement fluctuations
based upon the flux calibration of the stable calibrators,  
described below.  Using this standard deviation as the uncertainty in
measuring the flux of EC 53, the average peak brightness of 1230
mJy~beam$^{-1}$ in the last four epochs is $12 \sigma$ brighter than the brightness averaged over 
the first 6 epochs; the most enhanced 
peak brightness is $15 \sigma$ above the fainter level.

\begin{figure}[ht!]
\center
\includegraphics[width=0.45\textwidth]{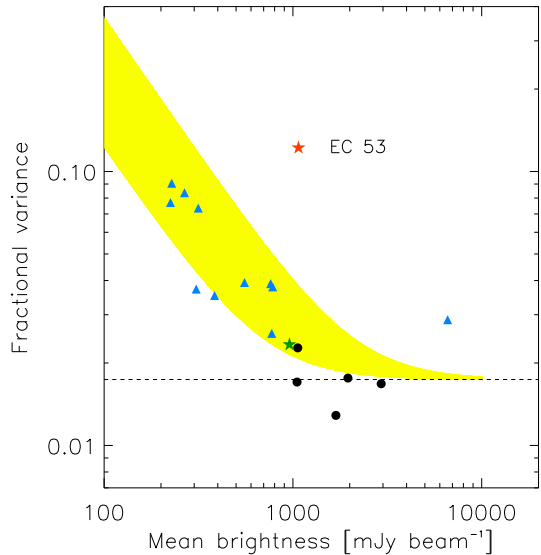}
\caption{
The fractional variance (the standard deviation of peak brightness over 11
epochs for each clump divided by the mean peak brightness for the corresponding clump) as
a function of the mean peak brightness. The black circles are flux calibrator
sources (Source 2, 3, 4, 6, and 7). The horizontal dashed line shows
the average fractional variance ($\sim$ 0.017) for the flux
calibrator sources, which corresponds to the calibration uncertainty. 
The red star indicates Source 5 (EC~53) which has a significant peak brightness variation
(the fractional variance $>$ 0.1). The green star is for the first 6
epochs for EC 53, over which the source was not observed to vary
(quiescent phase). While the data point of the red star symbol is obtained using all 11 epoch maps, the green star symbol is
obtained over the first six epochs (quiescent phase). The blue
triangles represent all other sources.  The lower bound and the upper bound of the yellow
shaded region correspond to the combination of the calibration uncertainty and the fractional variance estimated
from one and three times the mean rms noise (12 mJy~beam$^{-1}$) of the 11 maps, respectively.  
\label{fig:frac}}
\end{figure}

This brightness change is also highly
significant relative to the stability of all other bright sources in
the same field.
Figure \ref{fig:frac} shows the fractional variance (the
standard deviation of the peak brightness divided by the mean peak flux) as a function of the mean peak 
brightness for
each source.  
The average fractional variance of the five calibrators is $\sim 0.017\pm0.005$.
Because the noise in each map is $\sim 12$ mJy~beam$^{-1}$, the fractional
variance is higher for fainter sources.  Variations in the peak
brightness of faint sources are consistent with the combination of a calibration
uncertainty and
the fractional variance expected from the noise level of 12 mJy~beam$^{-1}$.

The fractional brightness changes of EC 53 stand out from all other
sources in Serpens Main  (Figure\ \ref{fig:frac}).  
The standard deviation of the peak brightness is much higher than expected from the uncertainty in 
 our flux calibration and the noise in the map.
However, in the first six epochs, EC 53 had a fractional variance of $\sim 2.5$\%,
consistent with expectations for a non-varying source.

The only other object that is inconsistent with the expected level of
variation is Source 1, the brightest source in the region, with a
standard deviation in peak brightness of $2.9$\%.  This variation is much smaller
than the flux change seen in EC 53 and requires a global evaluation of variability across all 8
regions to determine its significance (Mairs et al. submitted and Johnstone et
al. in prep). Source 1 is {\it not} included in the calibrator sample for
Serpens Main (see Figure\ \ref{fig:frac}).

\begin{figure*}[htp]
\centering
\includegraphics[width=.333\textwidth]{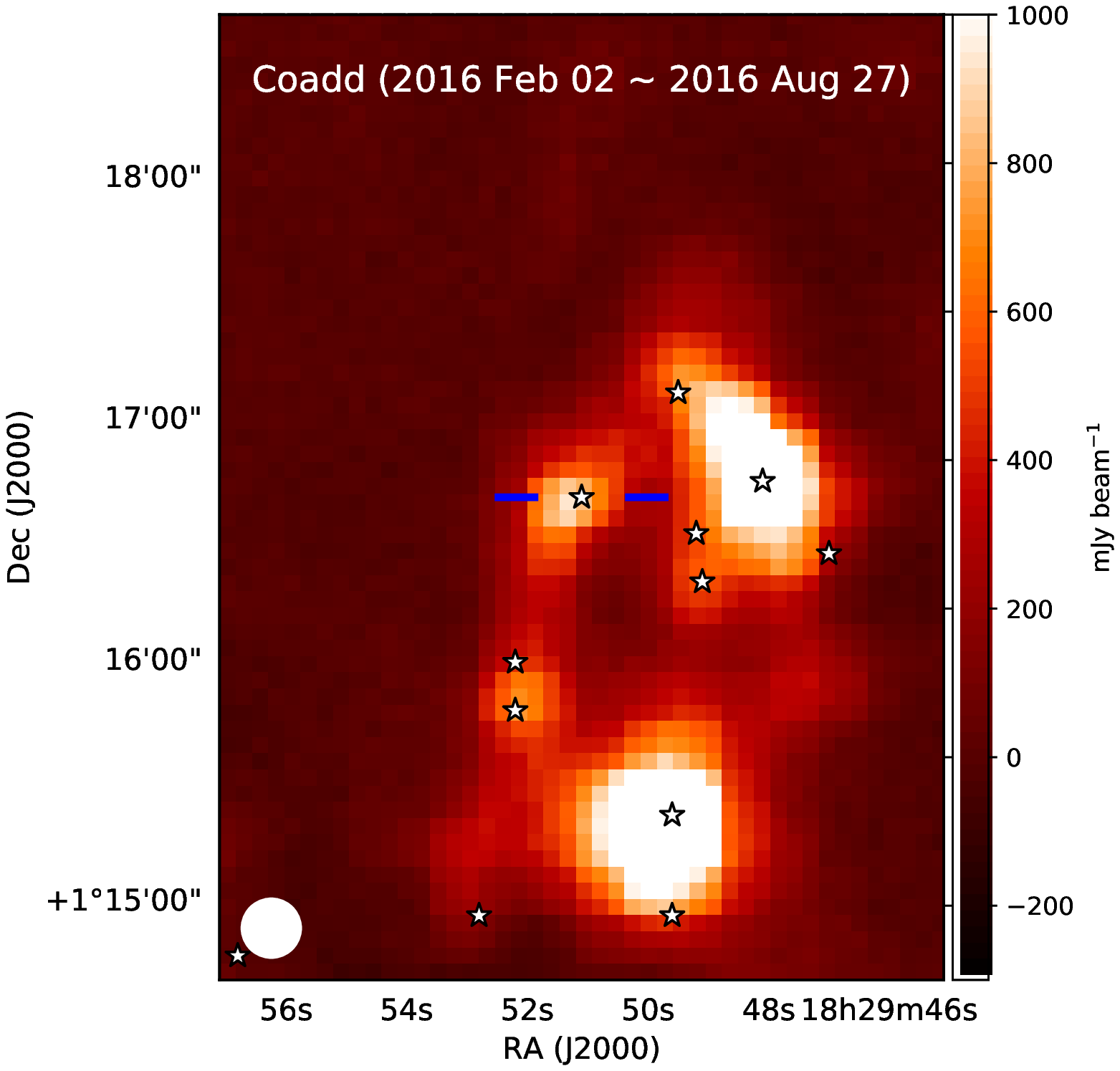}
\includegraphics[width=.333\textwidth]{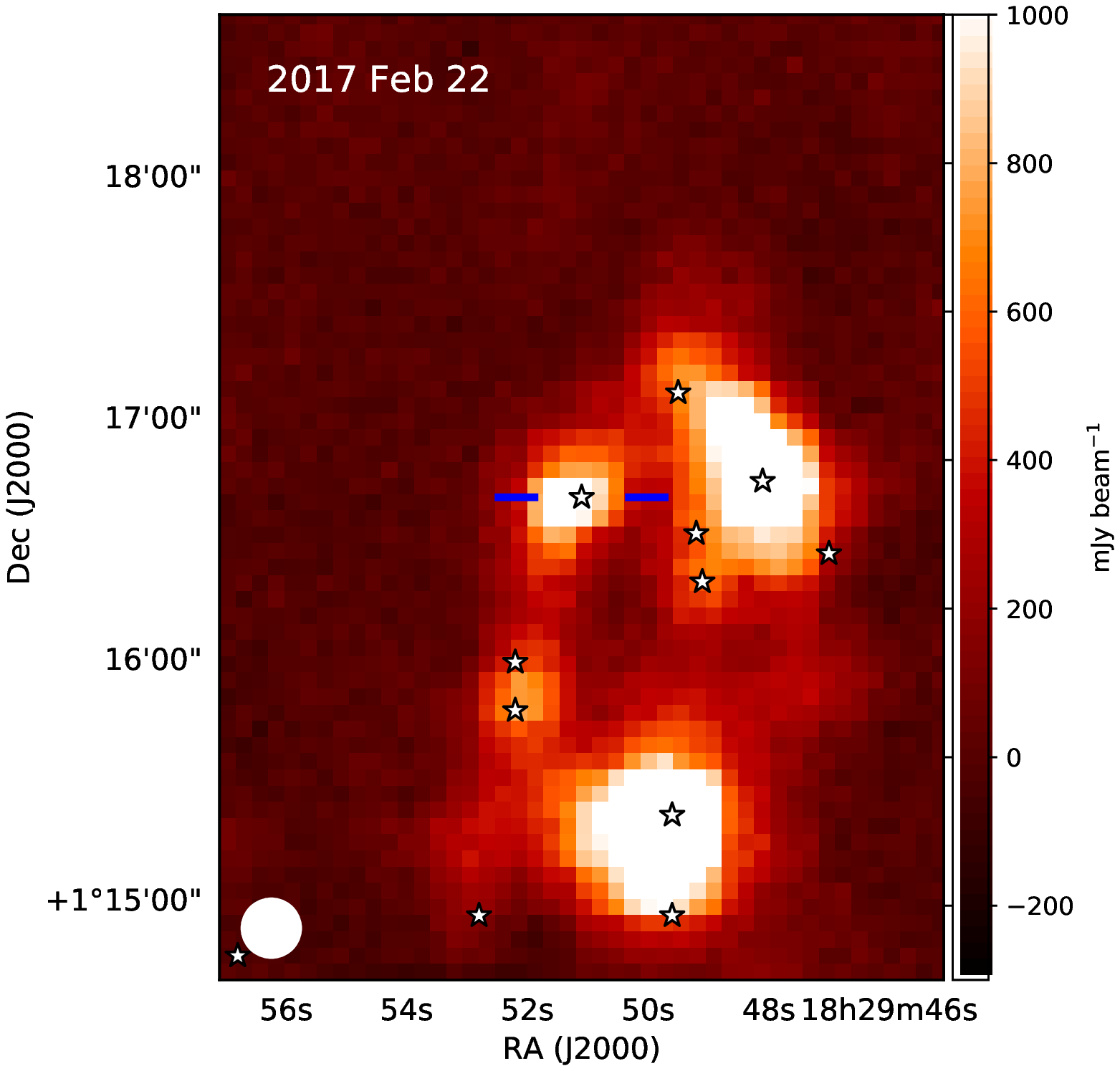}\\
\includegraphics[width=.333\textwidth]{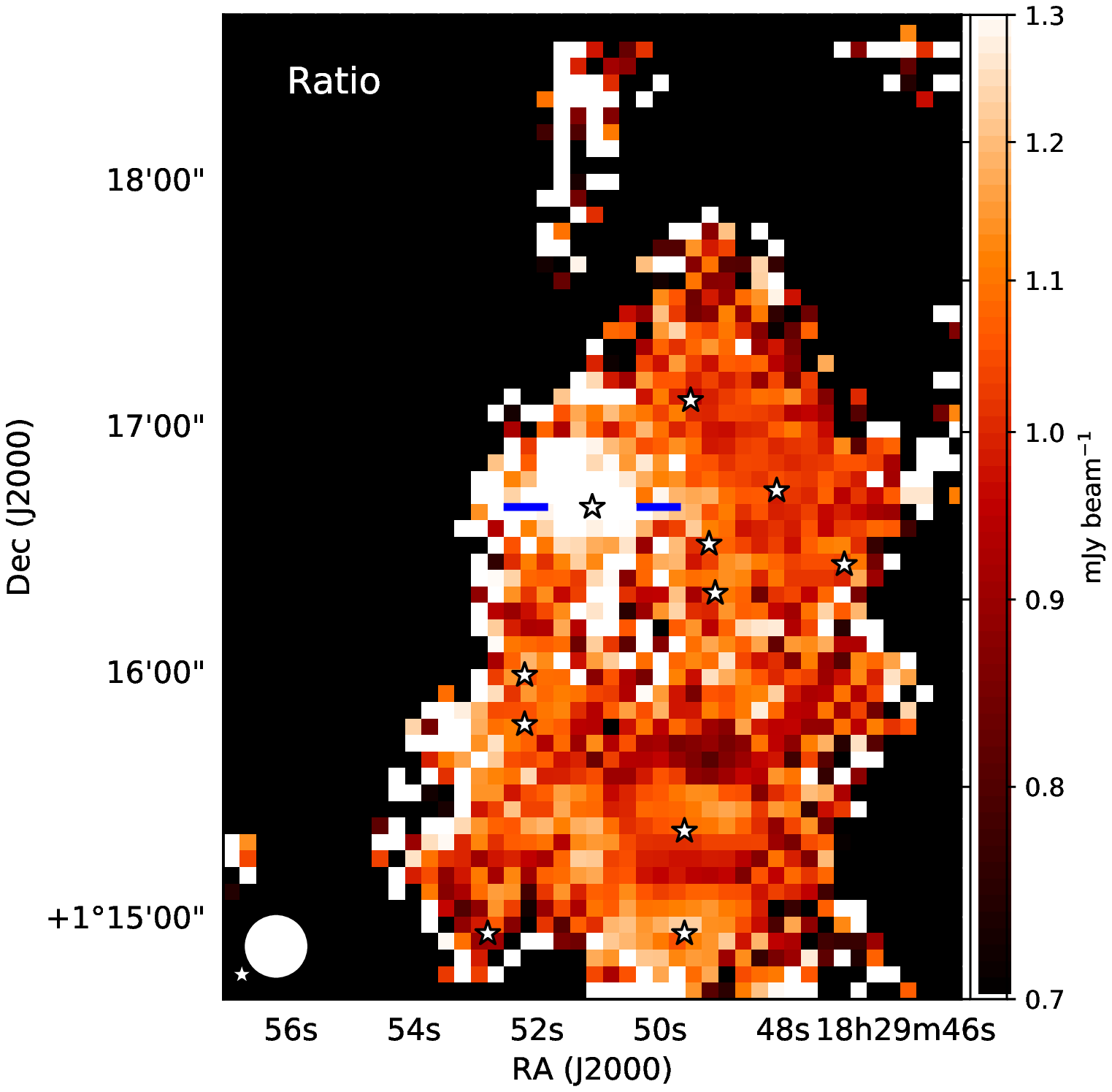}
\includegraphics[width=.333\textwidth]{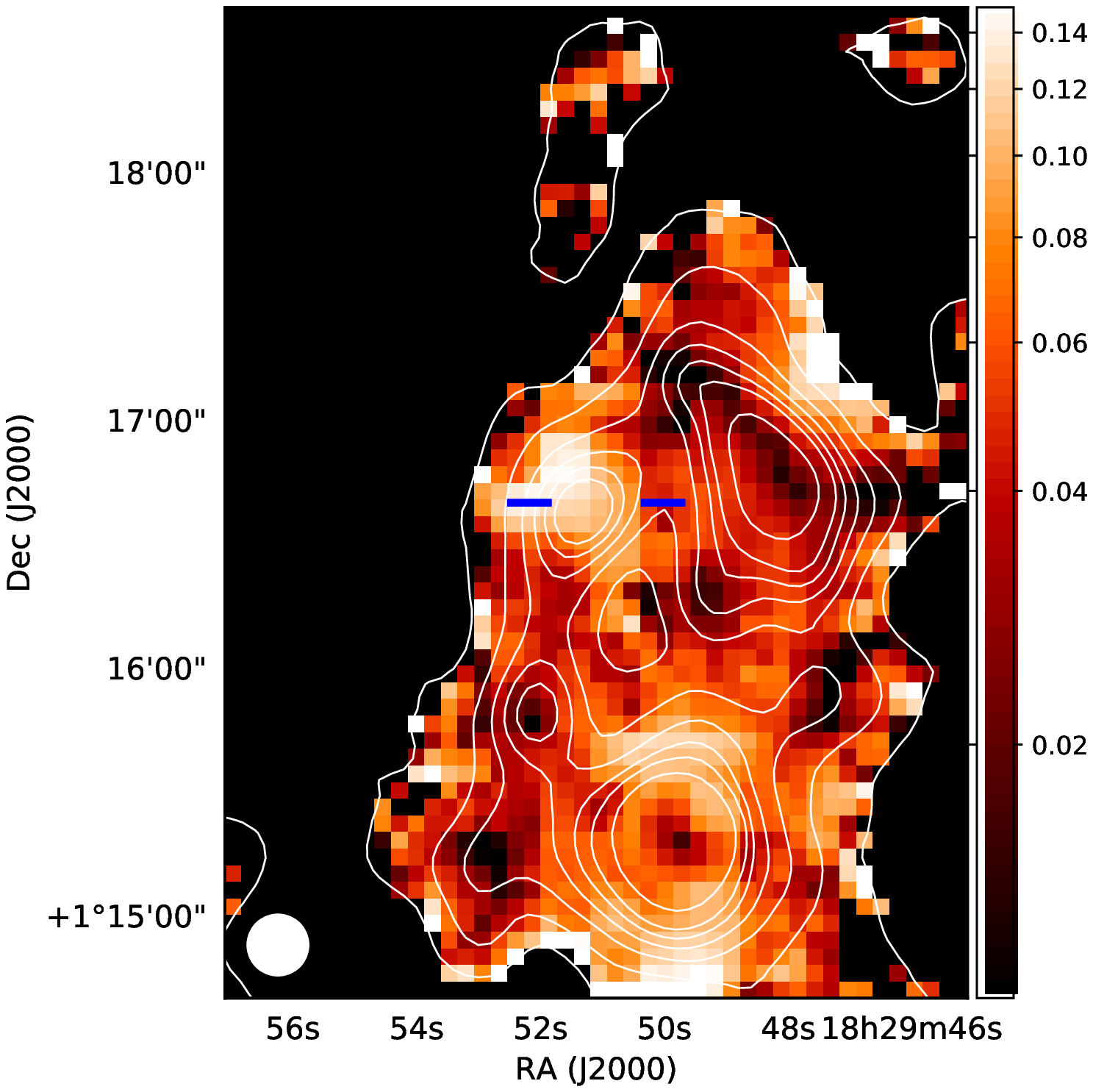}
\caption{SCUBA-2 850~$\mu m$ images. Upper left: A coadded map from
  23 February 2016 to 27 August 2016. Upper right: A map obtained on
  22 February 2017. The white stars mark the positions of protostars
  listed in \citet{dunham15}.  Lower left: The ratio of maximum phase (image in upper right panel) to quiescent phase (image in upper left panel). Lower right: The modified standard deviation map (see Section \ref{sec:result} for detailed description). Contours are from the coadded map with 5, 40, 80, 120, 160, 200, and 300$\sigma$ (1$\sigma$ $\sim$ 3.6 mJy~beam$^{-1}$). The blue horizontal lines mark the position of EC~53. The white filled circle in the bottom left corner in each image represents the JCMT beam at 850~$\mu m$. In the lower panels, pixels with intensity lower than 5$\sigma$ are masked.
\label{fig:ratio}}
\end{figure*}

The variability of EC 53 and the stability of other sources in the field is supported by comparing images before and after the flux jump.
In Figure \ref{fig:ratio}, EC 53 is fainter in the coadded 850~$\mu m$ image (upper left panel) obtained during the quiescent phase from 23 February 2016 to 27 August 2016) than in the image (upper right panel) of 22 February 2017. To evaluate variability throughout the image, we calculated the ratio between the two upper images 
(lower left panel) and also create a map of a modified standard deviation of every pixel (lower right panel). 
In the ratio image, the only brightness peak that stands out is EC 53. For the modified standard deviation image,  
all 11 images are first smoothed 
with a 6$\arcsec$ Gaussian beam to slightly suppress the noise between
pixels.  The standard deviation is then measured in each pixel, the average standard deviation in the image is subtracted, and then this background-subtracted standard deviation is divided by the average signal in each pixel.
The significance of variability of EC 53 is readily detected in this image of the modified standard deviation.
The other clumps have no signal at their centroid position, indicating that any fractional flux variations are small.  
The only other
locations on the map with high variability are in the wings of bright
sources, where small changes in the beam shape introduce large changes
between epochs.

\vskip 0.5cm
\section{Discussion} \label{sec:discussion}

\subsection{Lightcurves}

In this paper, we report the first detection of a sub-mm variable
source from our ongoing JCMT Transient survey.  In our monitoring, the
sub-mm flux of EC~53 was roughly constant in the first 6 months before
brightening in the 850 $\mu$m dust continuum by a factor of 1.5.
EC~53 is also variable in the near-IR, with $K$-band monitoring from 
1991--2011 indicating an amplitude change of $\sim 2$ mag with a
periodicity of $\sim 543$ days \citep{hodapp99,hodapp12}.   
During each period, the source brightened over about 110
days and then gradually faded over about 440 days \citep{hodapp99}.  

The right panel of Figure \ref{fig:lc} compares the sub-mm lightcurve
to the $K$-band lightcurve \citep{hodapp12}, extrapolated to the
epochs of our SCUBA-2 monitoring. The 850 $\mu$m lightcurve resembles
the $K$-band lightcurve, with brightness changes that are 4 times
smaller in the sub-mm (brightening by a factor of 1.5) than at
$K$-band (brightening by a factor of 6).  
The inset compares directly two lightcurves by overlaying each other; the 850 $\mu$m magnitude is multiplied by
4.  
During our observations, the maximum brightness of EC~53 occurred when
the field was behind the sun.  The maximum sub-mm brightness is estimated to be $\sim 1450$ mJy~beam$^{-1}$ (a 0.45 mag increase from the quiescent brightness of 960 mJy~beam$^{-1}$), by finding a crossing point of 
increasing and decreasing trends between the 6th and the 11th epoch.
The decreasing line is estimated by the linear least-square fitting of the last four data points (see the
lower right panel in Figure \ref{fig:lc}). 

\subsection{The SED}

In models of protostellar envelopes, sub-mm continuum emission is produced by dust that is heated by both the central protostar and the interstellar radiation field.  Any temporal changes in the sub-mm dust continuum emission are likely caused by a change in dust temperature in response to a change in the luminosity of the central object.  The total luminosity of the central object is produced by the stellar
photosphere and accretion, with any variability likely attributed to
accretion \citep[see also][]{hodapp12}.  In principle, the periodic near-IR lightcurves could also be
caused by a variable extinction \citep[e.g.][]{bouvier99,hamilton05,morales11}.
However, extinction would not affect the brightness at sub-mm wavelengths and can
therefore be ruled out.

To the first order, the change in luminosity of the central source warms every part of the envelope 
and this warming can be described by a power-law, $T \propto L^p$, where $p$ is $\frac{1}{4}$ for a black body. 
Therefore, we can make a rough assumption that the envelope warms up by a fixed fraction 
at every radial distance (see \citealt{johnstone13} for simple models) in the assumption of no variation 
in the opacity as temperature varies. As a result, the luminosity enhancement factor of the cold outer envelope
component can be considered as the enhancement factor of the central luminosity. 

\begin{figure*}[htp]
\centering
\includegraphics[width=.45\textwidth]{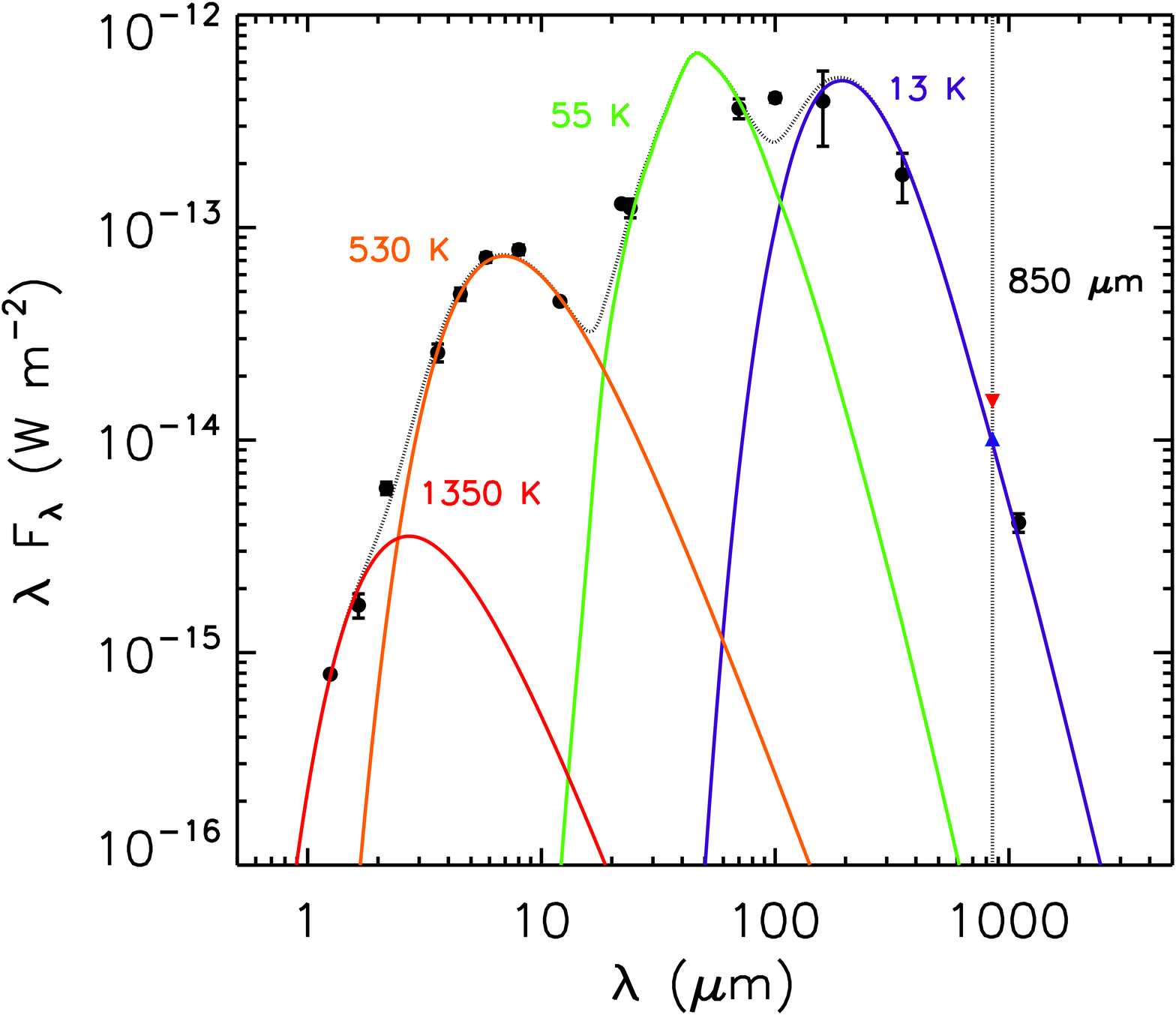}
\includegraphics[width=.45\textwidth]{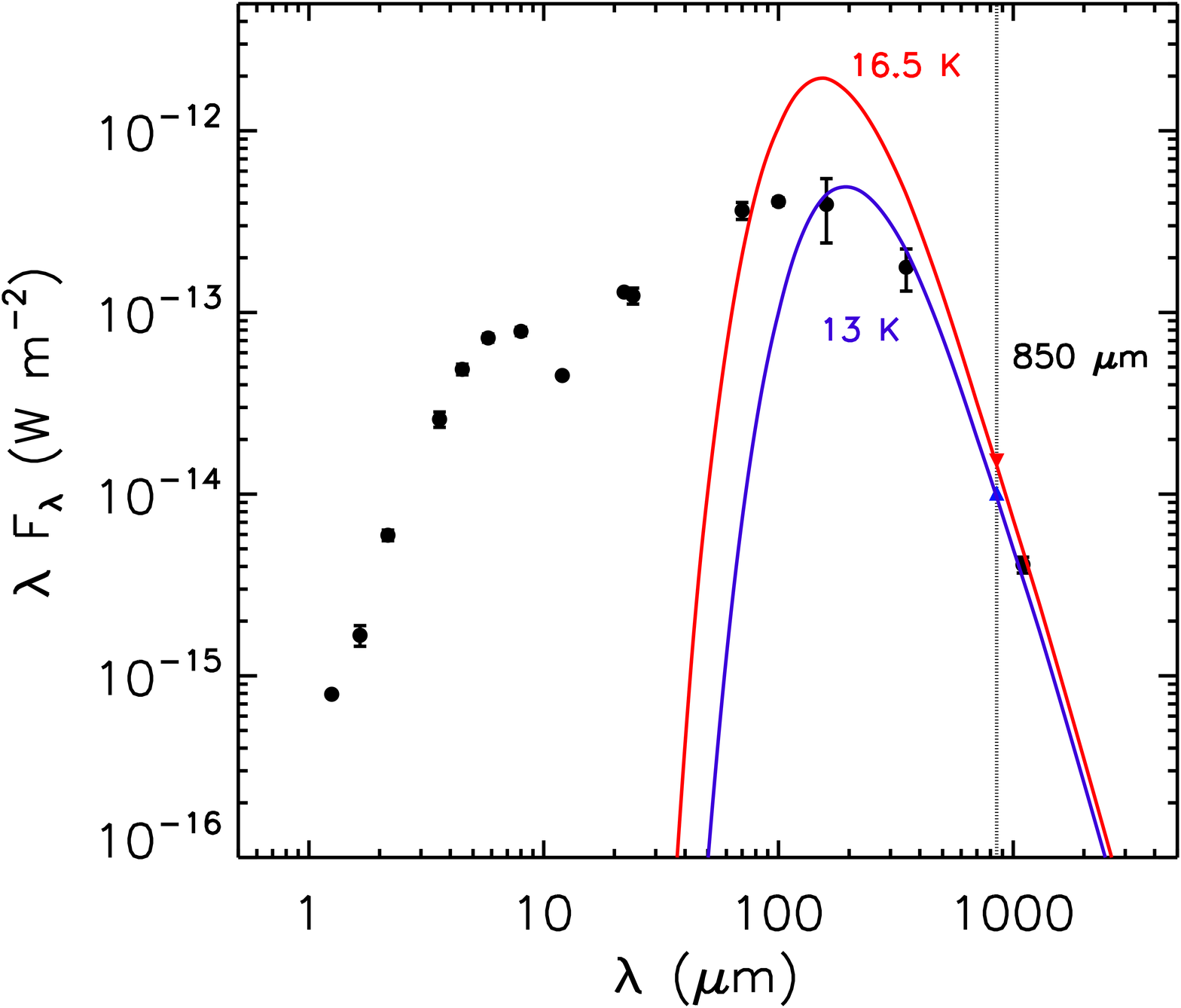}
\caption{Left panel: Filled circles show the spectral energy
  distribution of EC~53 at 1.25 $\mu$m to 70 $\mu m$  from \citet{dunham15}, at 100 $\mu$m and 160 $\mu$m from \citet{marton17}, at 350 $\mu$m (with an aperture size of 20\arcsec) from \citet{suresh16}, and at 850 $\mu$m from our data. 
  The 850 $\mu$m flux (blue triangle) is derived from the coadded map over the quiescent first 6 epochs. 
  The red upside-down triangle indicates the flux enhanced by a factor of 1.5, relative to the 
  quiescent phase, at 850 $\mu$m.
  The red and orange lines are the blackbody models, with temperatures of 1350~K and 530~K, which best fit the short wavelength observations. 
  The green and purple
  lines are the best-fit graybody models for the cold components at
  57~K and 18~K assuming a dust opacity of OH5  \citep{ossenkopf94}, respectively. The gray solid line shows 
  the sum of models for the each components. The vertical dashed line shows 850 $\mu m$. 
Right panel: the red line is the expected graybody model using OH5
dust opacity when the 850 $\mu m$ flux, which is mostly emitted from
the coldest component (T$_{0}$ = 13~K), is increased by a factor of 1.5
(T$_{Burst}$ = 16.5~K). Then, the luminosity increases by a factor of
$\sim$ 4.  
\label{fig:sed}}
\end{figure*}

In order to estimate the increase of (accretion) luminosity, we
assembled the spectral energy distribution
(SED) from \citet{cutri03} for 2MASS $JHK$ photometry, from
\citet{dunham15} for Spitzer photometry from 3.6 to 70 $\mu$m, from \citet{cutri13} for WISE 12 and 22 $\mu$m
photometry,
\citet{marton17} for Herschel/PACS 100 $\mu$m and 160 $\mu$m photometry, 
\citet{suresh16} for SHARC-II 350 $\mu$m, and our own 850 $\mu$m data.
The observed SED (Figure \ref{fig:sed}) is deconstructed into several
temperature components to evaluate what
change in luminosity could reproduce the significant variability seen in the sub-mm flux.  
This simple model does not consider any structures within the envelope and disk. The central source is likely
hotter than the hottest component (1350 K) found in the fitting (see the left panel of Figure \ref{fig:sed})
because it is deeply embedded and also masked by the very strong disk emission.

The 850 $\mu$m emission is mostly radiated from an outermost cold component,
which is fitted with a gray model with a dust temperature of 13~K
and a dust opacity of OH5 \citep{ossenkopf94}.
The cold outer dust component
absorbs much of the energy produced initially by the protostar, and
re-emits the energy predominantly at 
far-IR and submillimeter wavelengths.  If the luminosity of the
central object increases, the dust temperature in the outer envelope
should increase, thereby leading to brighter emission at 850 $\mu$m
\citep[see calculations by][]{johnstone13}. 
A flux enhancement of a factor of 1.5 at 850 $\mu$m
corresponds to an increase of the central luminosity by a factor of 4
and an envelope temperature by a factor of 1.3, respectively (see the right panel of Figure \ref{fig:sed}).
This is an extremely simple analysis but can still be useful for a rough estimate of the increase of the central luminosity.

The region observed to brighten at 850 $\mu$m is not 
spatially resolved and may include regions of the envelope that do not vary. The cold outer envelope is, at least in part, heated by the interstellar
radiation field \citep[e.g.][]{evans01, jorgensen06, dunham10, johnstone13} and is therefore less affected by changes in protostellar luminosity. Thus, the amplitude of any underlying sub-mm brightness variation produced by a change in protostellar luminosity might be diminished, leading to an underestimation of the amplitude of the protostellar variability itself. Moreover, our calculations assume that a constant fraction of the luminosity produced by the accreting protostar is absorbed by the outer envelope during both quiescence and burst.
A more sophisticated SED modeling with a continuum radiative transfer code, including the internal (accretion) 
and external (interstellar radiation field) heating sources, will be necessary for more detailed and realistic 
interpretation.

\subsection{Binarity}

The primary object in the system, EC 53 A, is responsible for the
measured change in the $K$-band emission \citep{hodapp12} and is therefore also
likely responsible for the change in sub-mm brightness.
\citet{hodapp12} speculated that EC 53 A is an unresolved, close
binary, perhaps formed by the scattering of the known companion, EC~53
B, third to a distant orbit (see \citealt{reipurth00} for a
description of the role of multiplicity in star formation).  Converting a periodicity in
accretion to binary properties is challenging.  A straightforward
interpretation would suggest that a 543-day period in accretion luminosity
with an asymmetric phased lightcurve corresponds to a binary with a 543-day period and an eccentric orbit.
However, the near-IR and sub-mm lightcurves of EC 53 appear
similar to the accretion bursts seen in hydrodynamical 
simulations of 
a circular, equal-mass binary surrounded by a circumbinary disk
\citep{munoz16}.  In these simulations, the accretion bursts occur
every fifth orbit, which would lead to an orbital period of $\sim 100$
days. 
This short period binary is hard to form directly via fragmentation of a molecular core or a protostellar disk.
The dynamical ejection of stars from unstable triple systems can
produce tight binary systems, even at very young ages \citep{moe17}.
If the accretion bursts are excited by a companion, the
companion could be stellar mass but could also be a giant planet, which are common at these periods.

\citet{bonnell92} initially suggested that 
FUor outbursts are induced by the tidal effect of the companion
object in a binary system.  At least some FUors are in binaries \citep[e.g.][]{koresko91,green16}.
The recent SMA observations toward 24
embedded protostars in the Perseus molecular cloud present a
chemical clue of burst accretion in an early evolutionary phase \citep{frimann17}; much
bigger CO sublimation sizes than expected from the current
protostellar luminosities can be explained 
by the timescale of the CO freeze-out onto grain surfaces being much longer than the dust cooling timescale \citep{lee07}. 
Although \citet{frimann17} found only a weak correlation between the presence of a disk and the past accretion
burst, three close binary systems in their sample show evidence of
accretion bursts occurring in the past. \citet{tofflemire16,tofflemire17} found definitive correlations between the binary orbit and periodic changes in accretion rate on two classical T Tauri stars.
Binary interactions may be an important mechanism of large and small burst versions of episodic accretion. 
In the EC~53 system, the relevant binarity would have to be unresolved, since the two resolvable components are separated by $\sim$
130~AU and would not induce a 543-day period.
Future observations with high resolution and high sensitivity
(e.g., with ALMA) toward EC~53 will enable us to understand the effect
of binary interactions on accretion bursts and the subsequent effects
on the dynamical and chemical evolution of the system.

The predictable timing of accretion bursts is nearly unique to EC 53 (see also L1634 IRS 7 by \citealt{hodapp15}
and LRLL54361 by \citealt{muzerolle13}).  
The classical FUor
eruptions last for decades and therefore cannot be repetitive on
timescales of individual humans.  EXor events have been seen to
repeat \citep[e.g.][]{aspin06,aspin10}, but unpredictably and
usually only after decades of quiescence. 
EC~53 is now returning to
quiescence.  According to the K-band lightcurve, 
we anticipate that EC~53 will continue to decay to a low state until February 2018
and will get brighter in May 2018, when Serpens Main is
observable in the night sky. 
Therefore, future observations of EC~53 provide the potential to
capture the moment of the accretion burst, to reveal the consequence of
an outburst on protostellar disk and envelope, and the connections between
mass accretion and mass ejection. Our Transient survey will continue to 
monitor EC~53 through the next maximum around May 2018 and until January 2019. 
We will be able to trace two phases of the 850 $\mu m$ lightcurve of EC~53
and pinpoint the best opportunities to catch the accretion burst.

\vskip 0.5cm
\section{Summary} \label{sec:summary}

We present the first detection of a variable source from the JCMT
Transient survey, which aims to evaluate sub-mm variability of
protostars in nearby low-mass star forming regions to find 
evidence of episodic accretion. EC 53 is one of the first sub-mm variables detected by any method, and the first variable protostar detected in a dedicated program.  In this paper, we:

\begin{enumerate}
\item{Identify and measure the peak brightness of 17 sub-mm sources in JCMT/SCUBA-2 850 $\mu$m images of
    Serpens Main, obtained in 11 epochs spread over 1.5 years.} 
\item{Detect a significant brightness increase by a factor of $\sim$1.3 with respect
    to the mean peak brightness at quiescent phase of the Class I YSO EC 53.  The maximum brightness at 850 $\mu$m is estimated to be $\sim 1.5$ times that at the faintest epochs, although this epoch of maximum brightness was unobservable because the field was behind the sun.} 
\item{Show that the 850 $\mu$ m lightcurve has a strong resemblance to the previously-known $K$-band lightcurve, which varies with a period of 543 days \citep{hodapp99,hodapp12}.}
\item{Calculate an increase of accretion luminosity by a factor
    of 4 from simple SED model fitting.}
\item{Describe the potential scientific utility that predictable changes in the accretion rate
provide and the importance of continued sub-mm monitoring of this source by the JCMT Transient Survey.}
\end{enumerate}



 

\acknowledgments
This paper used data from the JCMT Survey M16AL001 awarded by the East Asian Observatory.  
We thank the JCMT staff for carrying out the observations and contributions to the data reduction.
We greatly thank K. Hodapp for help with the ephemeris for EC~53.
We also thank S. Lee for help with the SED fitting.
Authors appreciate valuable comments by A. Pon, A. Scholz, H. Kirk, Y. Aikawa, and C. W. Lee.
This research was supported by the Basic Science Research Program
through the National Research Foundation of Korea (NRF) (grant
No. NRF-2015R1A2A2A01004769) and the Korea Astronomy and Space Science
Institute under the R\&D program (Project No. 2015-1-320-18)
supervised by Ministry of Science and ICT.   GJH
is supported by general grant 11473005 awarded by the National 
Science Foundation of China.  DJ is supported by the National Research Council of
Canada and by an NSERC Discovery Grant. SM is partically supported
by an NSERC graduate scholarship. 
MK was supported by by Basic Science Research Program through the
National Research Foundation of Korea(NRF) funded by Ministry of
Science and ICT (No. NRF-2015R1C1A1A01052160).
The JCMT is operated by the East
Asian Observatory on behalf of The National Astronomical Observatory
of Japan, Academia Sinica Institute of Astronomy and Astrophysics, the
Korea Astronomy and Space Science Institute, the National Astronomical
Observatories of China and the Chinese Academy of Sciences (Grant
No. XDB09000000), with additional funding support from the Science and
Technology Facilities Council of the United Kingdom and participating
universities in the United Kingdom and Canada.  
This research has made use of the Canadian Astronomy Data Centre operated by the National Research Council of Canada with the support of the Canadian Space Agency. This research used the services of the Canadian Advanced Network for Astronomy Research (CANFAR, specifically VOSpace), which in turn is supported by CANARIE, Compute Canada, University of Victoria, the National Research Council of Canada, and the Canadian Space Agency.

The authors wish to recognize and acknowledge the very significant cultural role
and reverence that the summit of Maunakea has always had within the
indigenous Hawaiian community.  We are most fortunate to have the
opportunity to conduct observations from this mountain.

%


\facility{JCMT (SCUBA-2) \citep{holland13}}
\software{Starlink \citep{currie14}, Astropy \citep{astropy}, Python version 2.7, APLpy \citep{aplpy}}

\end{document}